\documentstyle{agile7th}

\input epsf.sty
\input epsfig.sty
\input psfig.sty

\def \AAP #1 #2 {{\em Astron. Astrophys.\/} {\bf #1}, #2}
\def \AAL #1 #2 {{\em Astron. Astrophys. Lett.\/} {\bf #1}, L#2}
\def \AAR #1 #2 {{\em Astron. Astrophys. Rev.\/} {\bf #1}, #2}
\def \AAS #1 #2 {{\em Astron. Astrophys. Suppl. Ser.\/} {\bf #1}, #2}
\def \AJ #1 #2 {{\em Astron. J.\/} {\bf #1}, #2}
\def \ANNREV #1 #2 {{\em Ann. Rev. Astron. Astrophys.\/} {\bf #1}, #2}
\def \APJ #1 #2 {{\em Astrophys. J.\/} {\bf #1}, #2}
\def \APJL #1 #2 {{\em Astrophys. J. Lett.\/} {\bf #1}, L#2}
\def \APJS #1 #2 {{\em Astrophys. J. Suppl.\/} {\bf #1}, #2}
\def \APSS #1 #2 {{\em Astrophys. Space Sci.\/} {\bf #1}, #2}
\def \ASR #1 #2 {{\em Adv. Space Res.\/} {\bf #1}, #2}
\def \BAIC #1 #2 {{\em Bull. Astron. Inst. Czechosl.\/} {\bf #1}, #2}
\def \JSQRT #1 #2 {{\em J. Quant. Spectrosc. Radiat. Transfer\/} {\bf #1}, #2}
\def \MN #1 #2 {{\em Mon. Not. R. Astr. Soc.\/} {\bf #1}, #2}
\def \MEM #1 #2 {{\em Mem. R. Astr. Soc.\/} {\bf #1}, #2}
\def \PLR #1 #2 {{\em Phys. Lett. Rev.\/} {\bf #1}, #2}
\def \PASJ #1 #2 {{\em Publ. Astron. Soc. Japan\/} {\bf #1}, #2}
\def \PASP #1 #2 {{\em Publ. Astr. Soc. Pacific\/} {\bf #1}, #2}
\def \NAT #1 #2 {{\em Nature\/} {\bf #1}, #2}
\def \SAIT #1 #2 {{\em Mem.\ Soc.\ Astron.\ It.\/} {\bf #1}, #2}
\def \MESS #1 #2 {{\em The Messenger\/} {\bf #1}, #2}
\def \ASTRNACH #1 #2 {{\em Astron. Nach.\/} {\bf #1}, #2}
\def \AGPSR #1 #2 {{\em ASI Special Publication\/} {\bf #1}, #2}
\def \SCI #1 #2 {{\em Science\/} {\bf #1}, #2}

\begin{opening}

\title{The X-Ray/Radio/Flaring Properties of Cygnus X-3}
\author{M. L. McCollough$^{1}$, K. I. I. Koljonen$^{2}$, D. C. Hannikainen$^{2,3}$}
\institute{$^1$CXC/SAO/CfA, Cambridge, MA, U.S.A.\\
$^2$Aalto University Mets\"ahovi Radio Observatory, Kylm\"al\"a, Finland\\
$^3$Tuorla Observatory, Piikki\"o, Finland}
\date{} 
\end{opening}

\begin{document}

\oddpagefooter{}{}{} 
\evenpagefooter{}{}{} 
\medskip  

\begin{abstract} 

Cygnus X-3 is a unique microquasar.  Its X-ray emission shows a very strong
4.8-hour orbital modulation.  But its mass-donating companion is a Wolf-Rayet 
star.  Also unlike most other X-ray binaries Cygnus X-3 is relatively
bright in the radio virtually all of the time (the exceptions being the
quenched states).  Cygnus X-3 also undergoes giant radio outbursts (up
to 20 Jy).  In this presentation we discuss and review the flaring 
behavior of Cygnus X-3 and its various radio/X-ray states.  We present a revised
set of radio/X-ray states based on Cygnus X-3's hardness-intensity diagram 
(HID).  We also examine the connection of a certain type of activity to the 
reported AGILE/Fermi gamma-ray detections of Cygnus X-3.

\end{abstract}

\medskip

\section{Introduction}
Cygnus X-3 represents one of the most unusual X-ray binaries to have been 
observed (see Bonnet-Bidaud \& Chardin 1988 for a review).  It is at a distance
of $\sim$ 9 kpc (Predehl et al. 2000) in the galactic plane and is heavily obscured at 
optical wavelengths. Cygnus X-3 does not fit well into any of the established 
classes of X-ray binaries.  
It has a 4.8-hour orbital period, observed both in the X-ray (Parsignault et 
al. 1972) and the infrared (Mason et al. 1986), which is typical of a low mass 
X-ray binary. But infrared spectroscopic observations (van Kerkwijk et al. 1992,
Fender et al. 1999) indicate that the mass-donating star is a Wolf-Rayet 
star making the system a high mass X-ray binary.  In addition, 
Cygnus X-3 undergoes giant radio outbursts and there is strong evidence of 
jet-like structures moving away from Cygnus X-3 at 0.3--0.9c (Molnar et al. 
1988, Schalinski et al. 1995, Mioduszewski et al. 2001; see Fig. 1).  
The nature of the compact object is uncertain, but recent spectral studies 
(Szostek et al. 2008; hereafter SZM08 and Hjalmarsdotter et al. 2009) indicate 
that it may be a black hole.

\section{Discoveries from Multi-Wavelength Studies}

In studies (McCollough et al. 1997; 1999a, hereafter M99; 1999b) designed to 
classify Cygnus X-3's spectral states and behavior, the 20--100 keV hard X-ray 
(HXR) emission detected with {\it CGRO}/BATSE, was compared with the radio 
data from the Green Bank Interferometer (GBI; 2.25 and 8.3 GHz) and the Ryle 
Telescope (15 GHz), and with the 1--12 keV soft X-rays (SXR) detected with  
the {\it RXTE}/ASM.  Among the discoveries from these comparisons were:  

(a) During times of moderate radio brightness ($\sim 100$ mJy) and low 
variability, the HXR flux anti-correlates with the radio.  It is during this
time that the HXR reaches its highest level (see Fig. 2).

(b) During periods of flaring activity in the radio the HXR flux switches 
from an anti-correlation to a correlation with the radio.  In particular, for
major radio flares and the quenched radio emission (very low radio fluxes of 
10--20 mJy) which precedes them the correlation is strong (see Fig. 2).

(c) The HXR flux has been shown to anti-correlate with the SXR.  This occurs in
both the low and high SXR states (see Fig. 3).

(d) Results showing that flaring periods in the radio occur during the high SXR
states (Watanabe's 1994) were also confirmed. 

(e) It has been found from the {\it CGRO}/BATSE and {\it CGRO}/OSSE data 
that the spectrum of Cygnus X-3 (in the 20--100 keV band) becomes harder during 
times of radio flaring (McCollough 1999b).

\begin{figure}[ht!]
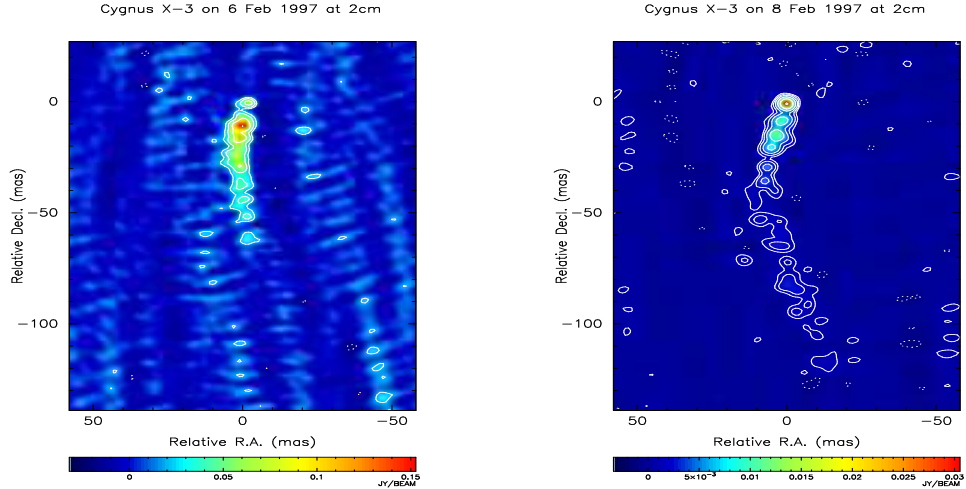

\centerline{\psfig{figure=mccollough_I_1fig1.ps,height=7cm,width=7.cm,angle=0}
\psfig{figure=mccollough_I_1fig2.ps,height=7cm,width=7.cm,angle=0}}
\caption{VLBI radio images obtained during a major radio flare of Cygnus X-3 (Mioduszewski 
et al. 2001).  The {\it left} panel shows a one-sided radio jet a few days 
after the peak in the radio.  The {\it right} panel shows how the jet has
evolved two days later.}
\end{figure}

\begin{figure}[ht!]
\centerline{\psfig{figure=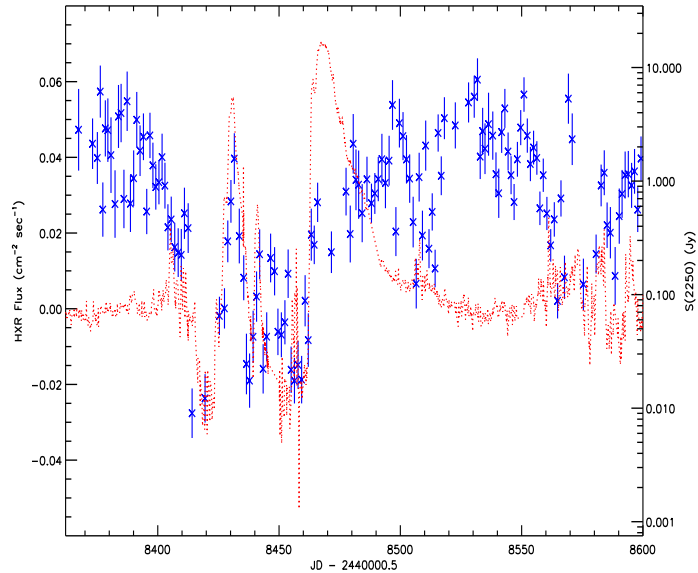,height=8cm,width=9.cm,angle=90}}
\caption{A comparison of GBI 2.25 GHz radio emission (red) with the {\it CGRO}/BATSE 
HXR emission (blue)
from M99.  Note the anti-correlation during times of radio
quiescence and the switch to a correlation during periods of quenched emission
and major radio flares.}
\end{figure}

\begin{figure}[ht!]
\centerline{\psfig{figure=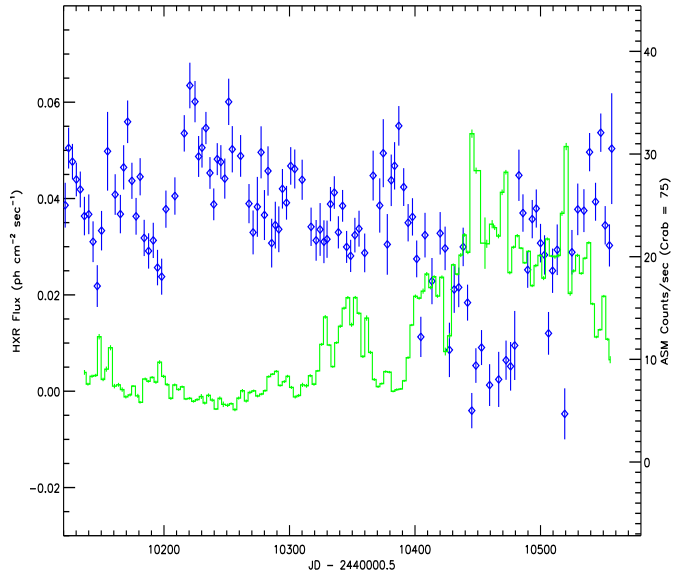,height=8cm,width=9.cm,angle=90}}
\caption{A comparison of  {\it RXTE}/ASM SXR emission (green solid line) with the 
{\it CGRO}/BATSE HXR
emission (blue diamonds) from McCollough et al. 1997.  Note the anti-correlation 
during all types of activity.}
\end{figure}

\section{Cygnus X-3 Radio/X-Ray States}

The X-ray spectra of Cygnus X-3 are notoriously complex (see Section 5 and Fig. 6).
Cygnus X-3 exhibits the canonical X-ray states seen in other X-ray
binaries (XRBs), 
namely the high/soft (HS) and low/hard (LH) states, in addition to the 
intermediate, very high and ultrasoft states (e.g. Szostek \& Zdziarski 2004, 
hereafter SZ04; Hjalmarsdotter et al. 2009). However, in Cygnus X-3, the strong 
radio emission is 
also classified into states of its own (Waltman et al. 1996; M99).  The X-ray 
emission has been found to be linked to radio emission in Cygnus X-3. M99 found 
that during periods of 
flaring activity in the radio the hard X-ray flux switches from an 
anti-correlation to a correlation with the radio. In addition, the HXR flux has 
been shown to anti-correlate with the soft X-rays in both canonical
X-ray states (McCollough et al. 1997).  Recently these X-ray and radio states 
were implemented into a more unified picture presented in SZM08:
the radio/X-ray states (see Fig. 4 and Table 1).

\begin{table}[ht]
\caption{Different Classification Methods of Cygnus X-3's States and Spectra (Koljonen et al. 2010)}
\begin{center}
\begin{tabular}{llllll}
\hline
\multicolumn{1}{c}{Canonical X-ray States} & \multicolumn{1}{c}{Radio States} & \multicolumn{1}{c}{SZ04 States} &  & \multicolumn{1}{c}{SZM08 States} & \multicolumn{1}{c}{New States}\\
\hline
& & group: & name:\\
& & & & & \\
&  & 1 & hard & quiescent & quiescent\\
low/hard & quiescent &  &  &  & \\
& & 2 & intermediate & minor flaring & transition\\
\\
intermediate & minor flaring & 3 & very high & suppressed &\\
& & & & & FHXR\\
& & & & ``post-flare" & \\
\\
&  & 4 & soft non-thermal & major flaring & FIM\\
& major flaring &  &  &  & \\
high/soft & & 5 & ultrasoft & quenched & FSXR\\
& quenched & & & & hypersoft\\
\hline
\end{tabular}
\end{center} 
\end{table}

\begin{figure}[ht!]
\centerline{\psfig{figure=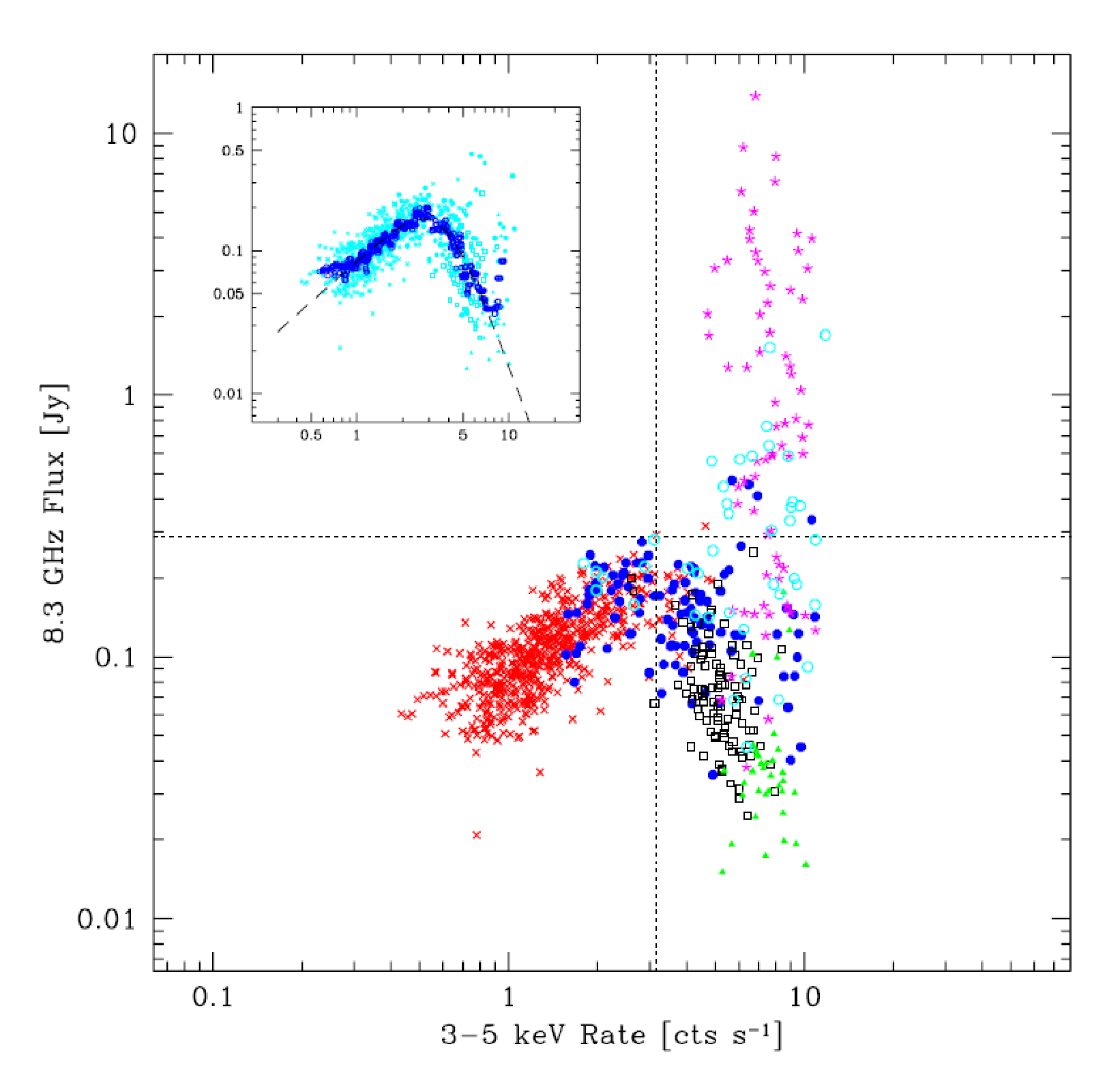,height=7cm,width=7.cm,angle=0}
\psfig{figure=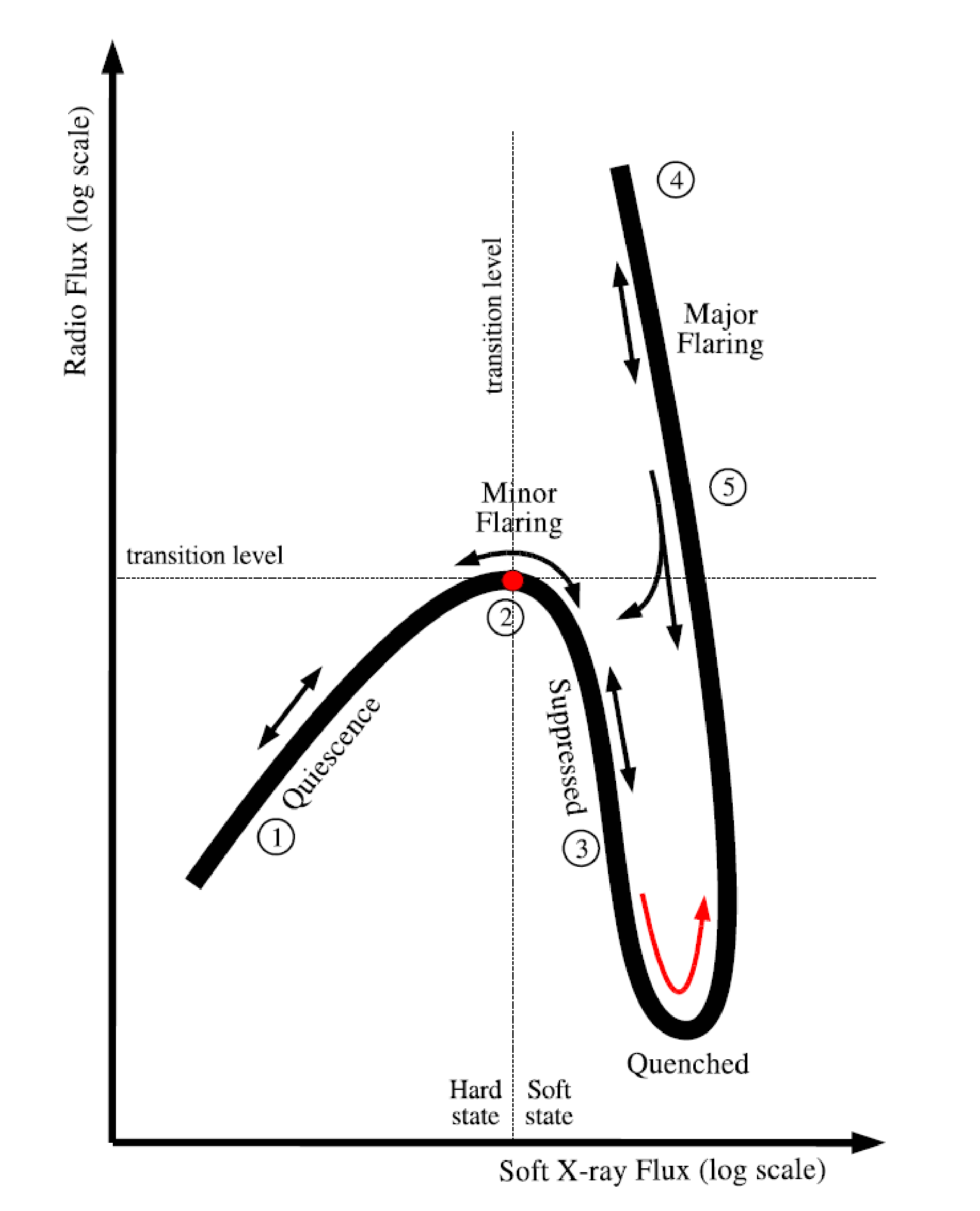,height=7cm,width=7.cm,angle=0}}
\caption{{\it LEFT:} X-ray ({\it RXTE}/ASM: 3--5 keV) vs. radio (GBI: 8.3 GHz) plot with with
different regions/states color coded (see SZM08 for full details).  {\it RIGHT:} A cartoon of the
X-ray - radio plot with states and paths between states labeled (SZM08).}
\end{figure}

\section{Cygnus X-3's Hardness-Intensity Diagram}

Following from previous work on black hole transients (Fender et al. 2004) a 
hardness-intensity diagram (HID) may yield insight into the nature of Cygnus 
X-3 despite it not being a transient in the strict sense of the word.  Thus,
using  {\it RXTE} pointed observations, we constructed a HID (Koljonen et al. 
2010) to study the behavior of Cygnus X-3.  The bands we used 
were chosen to probe different emission regions, with the SXR ($\rm 3-6~keV$) 
range representing the accretion disk and the HXR ($\rm 10-15~keV$) range representing 
the Comptonized part of the spectrum.  Given that Cygnus X-3 is also a persistent 
radio source we can add an additional dimension to the HID.  For each of the 
data points plotted we have used radio data from either GBI or the Ryle radio 
telescope to color code each data point by its radio flux density.  The 
resulting plot is shown in Fig. 5.

At first glance the plot appears surprisingly similar to other black hole XRB 
HIDs, but there are some very important differences. First of all, Cygnus X-3 
does not show hysteresis in the HID (Hjalmarsdotter et al. 2009) but simply 
increases in intensity as the spectrum softens. In black hole XRBs the LH state 
is present throughout the right branch, whereas in Cygnus X-3 the LH state is 
confined to the foot of the Q. The right branch in this case is dominated by 
X-ray flaring in Cygnus X-3. In addition, the flaring data are also spread into 
different hardnesses which wholly fill the inside area of the Q.

\begin{figure}[ht!]
\centerline{\psfig{figure=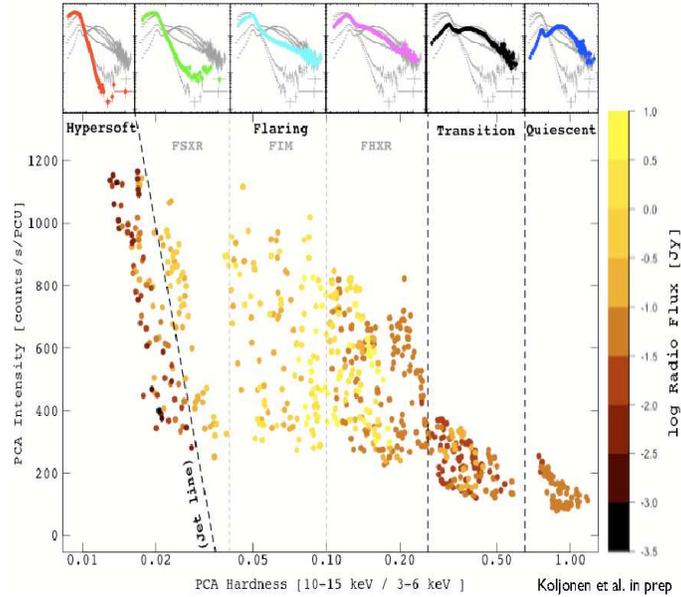,height=8cm,width=9.cm,angle=0}}
\caption{A HID for Cygnus X-3 created from pointed RXTE observations which also
have associated radio observations (Koljonen et al. 2010).  The color of the
data points represents the radio flux on a logarithmic scale.  Across the top of
the plot are average spectra for the labeled regions.}
\end{figure}

\section{Cygnus X-3's Spectra}

Cyg X-3 has a complex spectrum that can show drastic changes with change in
spectral state.  Among the spectral components are: (a) disk blackbody 
emission; (b) thermal Comptonized emission; (c) non-thermal 
Comptonized emission; (d) line emission; and (e) internal and 
external absorption.  Fig. 6 shows the succession of Cygnus X-3 spectra as it
progresses from a quiescent state (1), which is dominated by thermal Comptonized
emission, to a quenched state (6) in which the disk blackbody emission is
prominent.  The intermediate spectra have a strong non-thermal Comptonized 
emission component that appears as a power law tail to the spectrum and is most
likely tied into the major radio flares.

\begin{figure}[ht!]
\centerline{\psfig{figure=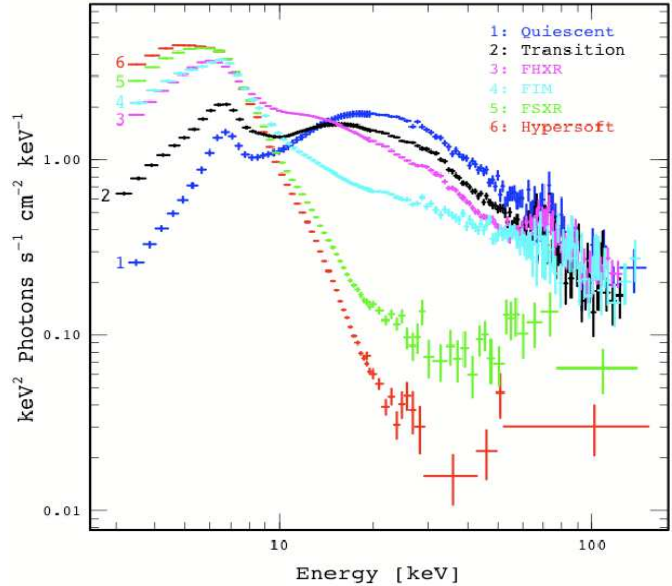,height=8cm,width=9.cm,angle=0}}
\caption{Plotted are typical {\it RXTE} X-ray spectra for Cygnus X-3 
(Koljonen et al. 2010).  They range from a
quiescent state (1-2) with a strong thermal Comptonized spectrum through the
various flaring states (3-5) with their non-thermal Comptonized spectral tail to
finally the hypersoft (6) spectrum whose emission is dominated by thermal disk
emission.  Note the spectral pivoting around 10 keV.}
\end{figure}

\section{New State Classification}

An examination of Cygnus X-3's HID allows us to create a revised set of 
radio/X-ray states based on
X-ray hardness and/or radio flux.  We have divided them into three distinct 
areas which we call quiescent, flaring and hypersoft.  

{\tt Quiescent:} The quiescent represents a region of moderate radio brightness,
low variability, and an X-ray spectrum which is dominated by thermal Comptonized
emission which peaks at around 20 keV and a prominent Fe line complex around 
6.7 keV.  We have divided this region into two sub-regions: {\it quiescent} and 
{\it transition}.  These sub-regions represent a gradual decrease and flattening 
of the hard part of the spectrum. 

{\tt Flaring:}  In turn the flaring state can be broken down into three 
sub-states according to X-ray hardness and the shape of the spectra: 
flaring/soft X-ray ({\it FSXR}), flaring/intermediate ({\it FIM}), and 
flaring/hard X-ray ({\it FHXR}).  The right-hand side of the {\it FHXR} is where
the minor flaring activity occurs.  From the center of the {\it FHXR}
through the {\it FIM} is the area where one finds the major radio flaring
activity.  Finally the {\it FSXR} represents an area where major radio flares
occur but the HXR flux has been greatly reduced.  This may indicate a type of
flare where there has been a strong interaction with Cygnus X-3's wind.

{\tt Hypersoft:} Finally there is the {\it hypersoft} state where a large 
fraction of the emission is from a disk blackbody and the non-thermal tail is 
an order of magnitude weaker (if present) than in the {\it FSXR}.  The radio
emission is highly suppressed (unless a previous major radio flare is still
decaying).  This is also what is known as the {\it quenched} state.  This state
is a precursor to the production of a major radio flare.

In Figs. 5 and 6 display the location in the HID and typical spectra for these
states.  In Table 1 the states and spectra of Cygnus X-3 from various studies 
have been listed.  It should be noted that as one goes down the
table there is a general transition in the behavior of Cygnus X-3.  But Cygnus 
X-3 can loop between states and also rapidly proceed between an upper state to 
a lower state while only spending minimal time in the intermediate states.

\section{The {\it Hypersoft} State and $\gamma$-Ray Emission}

During the quenched state, the radio emission falls to very low values ($\rm
1-20~mJy$), the HXR vanishes, and SXR reaches very high values.  As noted above
this is a precursor for the occurrence of major radio flares.   We have 
re-examined the X-ray spectra taken during quenching activity and shown that 
they can be divided into what we call {\it FSXR} (or 
ultrasoft of SZ04) and {\it hypersoft} state.  This 
{\it hypersoft} state has a power law tail which is an order of magnitude
fainter than that found in the quenched state of SZM08 (see Fig. 6 and 7).

Recent observations of $\gamma$-ray emission (Tavani et al. 2009, Abdo et al.
2009) from Cygnus X-3
have shown that the $\gamma$-ray emission appears to be directly related to the
activity associated with the 
{\it hypersoft} state and in turn the occurrence of major radio flares in Cygnus
X-3.

\begin{figure}[ht!]
\centerline{\psfig{figure=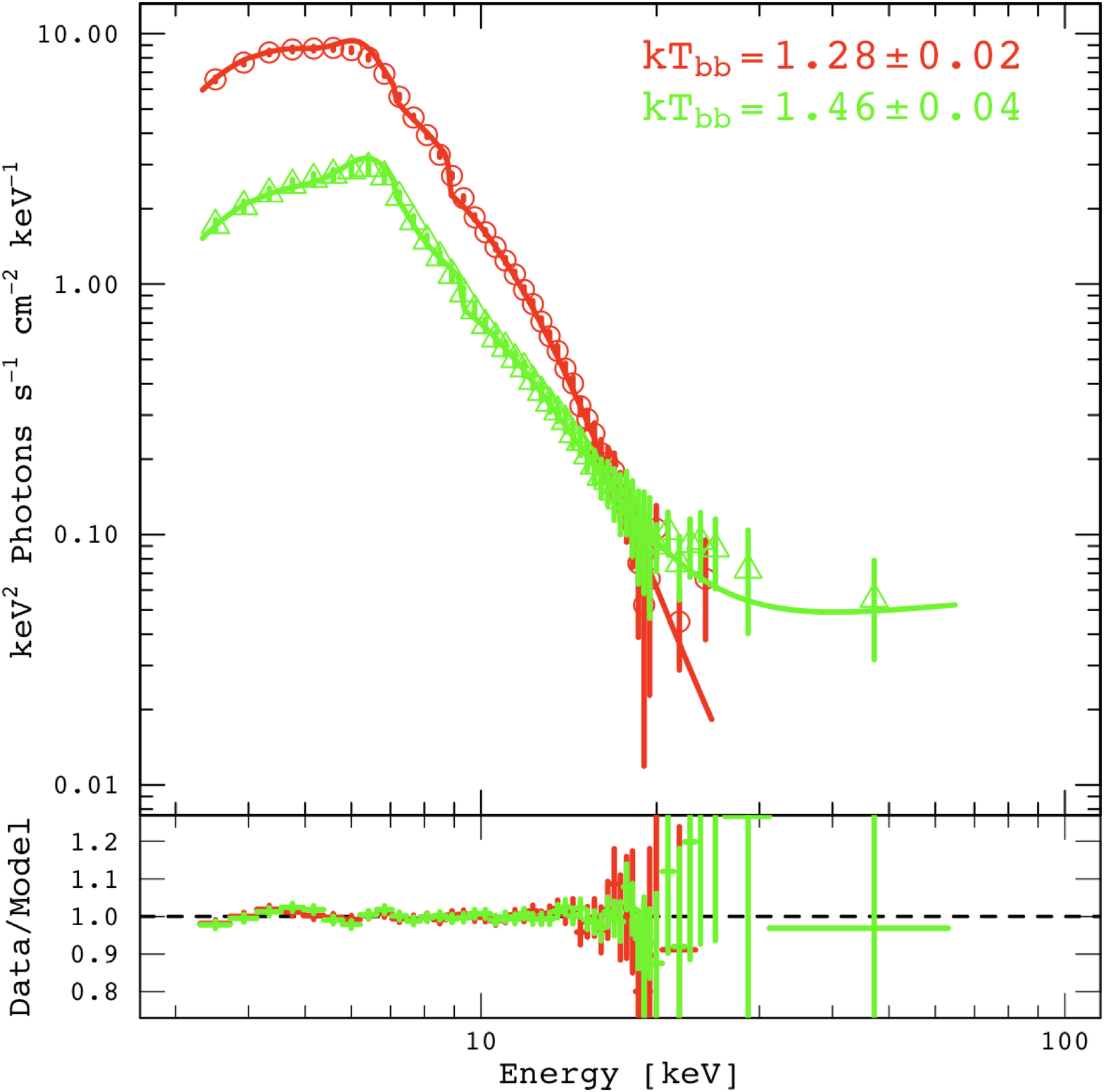,height=8cm,width=9.cm,angle=0}}
\caption{Plot are of a {\it FSXR} state (green open triangles) and a {\it hypersoft} 
state (red open circles) spectra along with their spectra fits.  Note that for the {\it hypersoft} 
state the hard X-ray tail and disk blackbody temperature is noticeably lower.}
\end{figure}

\section{Conclusions}

We have discussed the various correlations and radio/X-ray states of
Cyg X-3.  

From the construction of a unique HID for Cygnus X-3 with the radio flux as a
third dimension we have gained additional insight into the nature and
behavior of Cygnus X-3.  We have identified three broad states and several sub-states
which are delineated by intensity, X-ray hardness, and radio flux.  We have also
identified a ``new" very soft state that we call the {\it hypersoft} state.  This
state directly relates to the $\gamma$-ray emission  that has been observed from
Cyg X-3 (Tavani et al. 2009, Abdo et al. 2009).  This analysis lays the groundwork for a much
better understanding of the workings and nature of Cygnus X-3.

\section{Acknowledgments}

MLM wished to acknowledge support from NASA under grant/contract NNG06GE72G, 
NNX06AB94G and NAS8-03060.  KIIK gratefully acknowledges a grant from Jenny ja 
Antti Wihurin s\"a\"ati\"o and Academy of Finland grant (project num. 125189). 
DCH acknowledges Academy of Finland grant (project num. 212656).

\end{document}